%% file: main.tex
\documentclass[conference]{IEEEtran}
\IEEEoverridecommandlockouts

\usepackage{cite}
\usepackage{amsmath,amssymb,amsfonts}
\usepackage{algorithm}
\usepackage{algorithmic}
\usepackage{listings}
\usepackage{graphicx}
\usepackage{float}
\usepackage{dblfloatfix}
\usepackage{placeins}
\usepackage{textcomp}
\usepackage{pifont}
\usepackage[table]{xcolor}
\usepackage{tikz}
\usepackage{zref-savepos}
\usepackage{booktabs}
\usepackage{multirow}
\usepackage{makecell}
\usepackage{array}
\usepackage{url}
\usepackage{tabularx}
\usetikzlibrary{arrows.meta,positioning,fit,backgrounds,calc,shapes.geometric,decorations.pathreplacing}
\usepackage[most]{tcolorbox}
\usepackage{hyperref}

\def\BibTeX{{\rm B\kern-.05em{\sc i\kern-.025em b}\kern-.08em
    T\kern-.1667em\lower.7ex\hbox{E}\kern-.125emX}}

\begin{document}

\title{QiMeng-TheoremV: Agentic Theorem Proving with Speculated IRs for Large-Scale RTL Formal Verification}

\author{\IEEEauthorblockN{Chenrui Cao\textsuperscript{1,2}, Chongxiao Li\textsuperscript{2},
Yutong Wu\textsuperscript{2,3}, Hongqin Lyu\textsuperscript{2,3}, Di Huang\textsuperscript{2,3},\\
Chuanhao Wang\textsuperscript{3}, Lei Zhang\textsuperscript{4}, Chao Wang\textsuperscript{1},
Tian Zhi\textsuperscript{2,3}, Rui Zhang\textsuperscript{2,3},\\
Zidong Du\textsuperscript{2,3}, Qi Guo\textsuperscript{2,3}, Xing Hu\textsuperscript{2,3*}, Yunji Chen\textsuperscript{2,5,3}}\\
\IEEEauthorblockA{\textsuperscript{1}University of Science and Technology of China, Hefei, China\\
\textsuperscript{2}State Key Lab of Processors, Institute of Computing Technology, Chinese Academy of Sciences, Beijing, China\\
\textsuperscript{3}University of Chinese Academy of Sciences, Beijing, China\\
\textsuperscript{4}Cambricon Technologies, Beijing, China\\
\textsuperscript{5}Institute of AI for Industries, Chinese Academy of Sciences, Nanjing, China
}

\thanks{*Corresponding author: Xing Hu (\mbox{huxing@ict.ac.cn}). First author contact: \mbox{caochenrui@mail.ustc.edu.cn}.}
}

\maketitle

\begin{abstract}

As LLMs speed up automated Register-Transfer Level (RTL) code generation, verifying that generated designs meet their specs has become a key bottleneck. We present TheoremV, an agentic framework that uses Lean~4 theorem proving to automatically build complete proofs of end-to-end functional equivalence between large-scale RTL designs and high-level reference models. TheoremV bridges two gaps: the abstraction gap between cycle-accurate RTL and high-level reference models, and the representation gap between RTL and proof languages. For the abstraction gap, TheoremV introduces Speculated IRs: an LLM agent proposes candidate intermediate models, quickly filters out bad ones via differential simulation, and revises them using proof feedback---breaking an end-to-end proof across abstraction levels into a chain of equivalence lemmas. For the representation gap, a Yosys-based RTL-to-Lean compiler keeps industrial frontend coverage while preserving word-level signals and module hierarchy, and port-level dependency analysis removes false combinational loops at module interfaces.

On tasks where commercial high-level equivalence checkers time out and existing automated provers fail, TheoremV proves the functional equivalence of silicon-proven floating-point and matrix units (XiangShan Fudian and NVDLA CMAC) against high-level reference models, as well as the PicoRV32 core against the official RISC-V Sail spec---finding a real RISC-V spec bug (in the FENCE instruction) in PicoRV32.

\end{abstract}


\input{sections/0_introduction}
\input{sections/1_related_work}
\input{sections/2_methods}
\input{sections/4_evaluations}
\input{sections/99_conclusion}

\input{sections/100_AI}

\bibliographystyle{IEEEtran}
\IEEEtriggeratref{23}
\bibliography{references}

\end{document}

%% file: sections/0_introduction.tex
\section{Introduction}

\begin{table}[!tbp]
\caption{Comparing TheoremV with recent RTL theorem proving work.}
\label{tab:compare}
\centering
\scriptsize
\setlength{\tabcolsep}{1pt}
\renewcommand{\arraystretch}{1.05}
\renewcommand{\tabularxcolumn}[1]{>{\centering\arraybackslash\linespread{0.9}\selectfont}m{#1}}

\newcolumntype{C}{>{\hsize=0.945\hsize}X}
\newcolumntype{H}{>{\hsize=1.22\hsize\columncolor{blue!6}}X}

\begin{tabularx}{\columnwidth}{@{} >{\bfseries\raggedright\arraybackslash}m{1.92cm} C C C C H @{}}
\toprule
Aspect & Chicala \cite{chicala} & Rtl2lean \cite{rtl2lean-paper} & Circuit-Prover \cite{circuitprover} & Infini-Verif \cite{infiniverif} & \textbf{TheoremV (Ours)} \\
\midrule
Frontend & Chisel & PySlang & Chisel & CIRCT & \textbf{Yosys} \\
\specialrule{0.2pt}{0.3pt}{0.3pt}
Syntax cov. & Narrow & Narrow & Narrow & Medium & Large \\
\specialrule{0.2pt}{0.3pt}{0.3pt}
End-to-end equiv. & \ding{55} & \ding{55} & \ding{55} & \ding{55} & \textbf{\checkmark} \\
\specialrule{0.2pt}{0.3pt}{0.3pt}
Automated & \ding{55} & \checkmark & \checkmark & \checkmark & \textbf{\checkmark} \\
\specialrule{0.2pt}{0.3pt}{0.3pt}
Abstraction & Manual inv. & 4-tier split & Lemma reuse & Circuit semantic lifting & \textbf{Speculated IRs} \\
\specialrule{0.2pt}{0.3pt}{0.3pt}
Benchmarks & Xiangshan / Rocket Mul/Div & 6 RTL designs from Opencores & 63 circuits, Xiangshan / Rocket Mul/Div & MIPS, Ibex, Spatz & \textbf{Fudian, CMAC, PicoRV32} \\
\bottomrule
\end{tabularx}
\end{table}

Advanced large language models (LLMs) have excelled at automated register-transfer-level (RTL) code generation for integrated circuit (IC) design~\cite{revisiting-verilogeval}. To apply the large volume of generated RTL code in industrial scenarios, ensuring its correctness is of critical importance. Even after extensive UVM testing, rare error cases may still remain uncovered. For example, the Intel Pentium floating-point division bug occurred in only about one out of nine billion random operand pairs, thereby escaping simulation-based verification and  leading to a costly recall~\cite{harrison-fdiv}. Consequently, \textbf{formal verification}—which guarantees correctness for all legal inputs—has become one of the ultimate safeguards for hardware verification.

However, formal verification of large-scale hardware designs remains challenging. 
Model checking and equivalence checking are widely used formal methods for circuit verification, but both suffer from state-space explosion as system complexity grows.
Theorem proving avoids state-space explosion by employing abstraction, induction, and modular reasoning to construct proofs of correctness without enumerating every state; however, it demands substantial human effort and expertise to build formal models, discover invariants, and write proofs. 

Recent progress in LLM-assisted mathematical theorem proving~\cite{alphaproof} has inspired experimental efforts to automate hardware proof generation~\cite{rtl2lean-paper,circuitprover,infiniverif}. Nevertheless, these efforts still face two limitations.
\ding{182}~\textbf{Limited verification scope.}
Existing work focuses largely on local properties, whereas practical hardware development requires end-to-end functional equivalence between RTL and
independently developed high-level reference models.
\ding{183}~\textbf{Limited circuit scale.} Existing methods mainly target standalone modules or small arithmetic blocks, falling short of complex designs.

Automated theorem proving for large-scale circuits is fundamentally hard due to the huge gap between the Verilog design under verification (DUV) and the high-level reference model (REF). This gap severely impedes automatic invariant discovery and causes the state space and proof paths to explode. 
We therefore propose \textbf{TheoremV}, a validation framework for automatically constructing complete, machine-checkable Lean proofs of \textbf{end-to-end} functional equivalence between large-scale RTL designs and high-level reference models. As shown in Table~\ref{tab:compare}, TheoremV supports broad Verilog syntax, establishes end-to-end equivalence against reference models, and scales to complex circuits by bridging two gaps:

\ding{182}~\textbf{The abstraction gap between RTL and reference models.}
RTL implements functionality through parallel datapaths and state transitions across multiple
clock cycles, while reference models are typically implemented at a much higher level of abstraction.
Their intermediate states and execution steps may differ significantly; therefore, structural decomposition that splits hardware proofs along module boundaries or execution stages~\cite{rtl2lean-paper,circuitprover} does not resolve the difficulty of proving equivalence, as shown in Fig.~\ref{fig:fma_decomposition}(b). 
Our central \textbf{insight} is that introducing intermediate models between RTL and its reference model decomposes an equivalence proof spanning
multiple abstraction levels into smaller proofs between adjacent models. 
For example, connecting the RTL implementation of a floating-point FMA unit to the SoftFloat C
reference model~\cite{softfloat} can be decomposed into smaller equivalence proofs through a formal specification of IEEE~754 FMA.
To automate this decomposition, TheoremV introduces an agentic workflow based on \textbf{Speculated IRs} that automatically \textit{proposes, filters, and revises} tentative
intermediate reference models to reduce the proof complexity.

\ding{183}~\textbf{The representation gap between RTL and the proof language.}
Translating RTL into Lean requires faithfully capturing parallel behavior and cycle-level
semantics while preserving module hierarchy to support definition reuse and avoid the excessive
model growth caused by full flattening. TheoremV designs an RTL-to-Lean compiler that integrates closely with existing EDA flows.
Built on Yosys's RTL frontend and RTLIL representation~\cite{yosys}, the RTL-to-Lean compiler inherits
the frontend's syntax coverage while preserving word-level signals and module hierarchy.
This allows module instances to share definitions, avoiding the code growth caused by fully
expanding each instance.
Our port-level dependency analysis resolves false combinational loops at module interfaces,
enabling the generation of executable, purely functional Lean state machines.

TheoremV handles verification tasks substantially larger than those supported by conventional
formal verification tools and prior automated theorem-proving methods. 

Our main contributions are summarized as follows:
\begin{itemize}
    \item \textbf{Speculated IR-Driven Proof Decomposition.} We introduce an agentic workflow that dynamically proposes, simulation-filters, and revises executable \textit{Speculated IRs} to decompose multi-level equivalence proofs into tractable sub-goals without manual theorem libraries.
    \item \textbf{Hierarchical RTL-to-Lean Compilation.} We build a Yosys/RTLIL-based compiler that preserves word-level semantics and module hierarchy while resolving false combinational loops into purely functional Lean state machines.
    \item \textbf{Industrial Evaluation and Bug Discovery.} TheoremV proves end-to-end equivalence for taped-out floating-point and matrix arithmetic units against high-level reference models, and uncovers a specification-violating defect in the \texttt{FENCE} instruction of PicoRV32~\cite{picorv32}.
\end{itemize}

%% file: sections/1_related_work.tex
\section{Related Work}
\label{sec:related}

\paragraph{Hardware Model Checking}
Open-source (SymbiYosys~\cite{yosys}, riscv-formal~\cite{riscvformal}) and commercial (JasperGold, VC Formal) model checkers excel at local properties, while CIll~\cite{cill} uses LLMs to synthesize inductive invariants from IC3/PDR counterexamples for PicoRV32. However, on complex arithmetic datapaths, model checking still suffers from state-space explosion and often degrades into bounded checks. In contrast, our work uses algebraic abstraction and induction in Lean~4 to prove correctness over all inputs and unbounded cycles.

\input{figs/fig-decomposition-v2}

\paragraph{Hardware Semantics in Theorem Provers}
Formalizing hardware in proof assistants follows three paths. First, embedded HDLs like Kami~\cite{kami} (Coq), Sparkle~\cite{sparkle} (Lean~4), and CktFormalizer~\cite{cktformalizer} require rewriting designs and cannot verify existing RTL. Second, direct Verilog semantics define formal execution rules, such as Lutsig~\cite{loow2022} via topological sorting and Choi et al.~\cite{revamping-verilog} via least fixed points. Third, frontend translators target narrow subsets, such as Rtl2lean~\cite{rtl2lean-paper} (SystemVerilog), CircuitProver~\cite{circuitprover} and Chicala~\cite{chicala} (Chisel), and InfiniVerif~\cite{infiniverif} (CIRCT/MLIR). Unlike custom frontends, TheoremV builds on Yosys for broad RTL support and emits clean, executable Lean state machines for fast testing and proving.

\paragraph{Formal Verification of Floating-Point Units and Processors}
Verifying floating-point units and processors has long required heavy manual effort. Classic milestones, including Harrison's IA-64 division proof~\cite{harrison-fdiv}, Kaivola et al.'s Pentium~4 multiplier proof~\cite{kaivola}, and Flocq~\cite{flocq}, either checked high-level algorithms or took months to years of manual work. For processors, Choi et al.~\cite{revamping-verilog} manually proved a pipelined RISC-V core, while Lee and Kang~\cite{ooo-multiprocessor} verified an out-of-order multiprocessor with LLM help yet still needed heavy guidance. Our framework automates this workflow, verifying production RTL like XiangShan Fudian and PicoRV32 directly against SoftFloat and RISC-V Sail~\cite{sail} models.

\paragraph{LLM Theorem Proving and Agentic Hardware Verification}
Inspired by closed-loop math provers such as DSP~\cite{dsp}, LEGO-Prover~\cite{legoprover}, DeepSeek-Prover~\cite{deepseekprover}, and AlphaProof~\cite{alphaproof,alphaproof-nexus}, recent tools extend this paradigm to RTL: Chicala~\cite{chicala} checks Chisel in Stainless with manual invariants; Rtl2lean~\cite{rtl2lean-paper} splits proofs by modules and fixed cycles; CircuitProver~\cite{circuitprover} reuses sub-block lemmas; and InfiniVerif~\cite{infiniverif} lifts loops to check local contracts. Yet when linking multi-cycle RTL to external C or ISA golden models, structural splitting along modules or cycles breaks down. Instead, our agents build and test executable \textbf{Speculated IRs} to bridge this gap.

%% file: figs/fig-decomposition-v2.tex
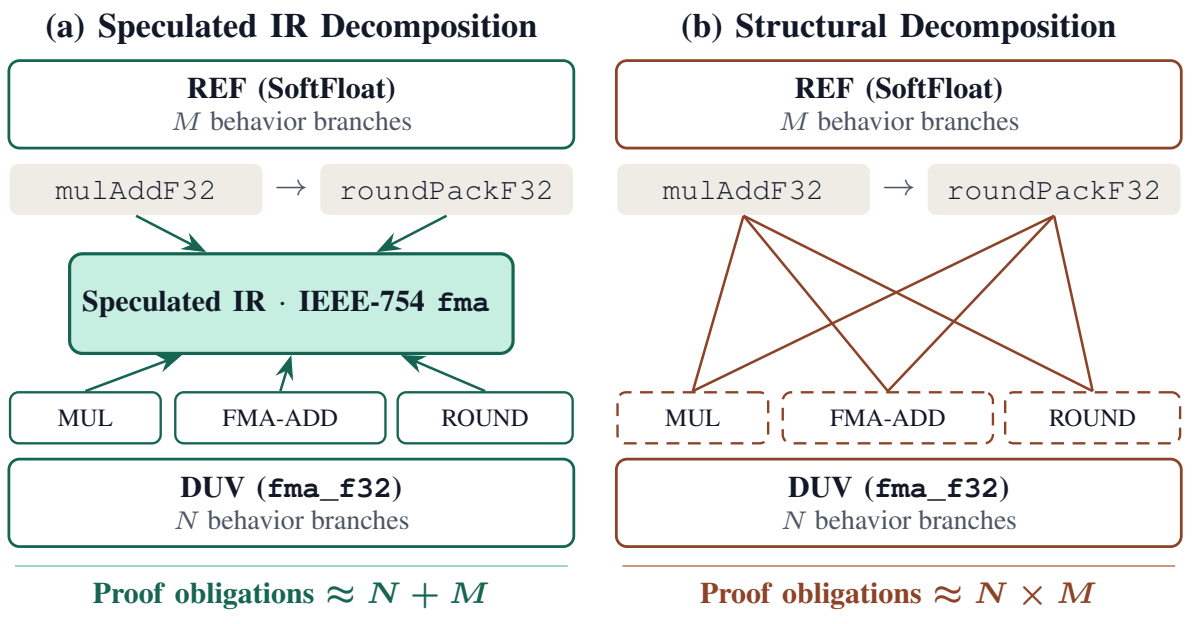
\begin{figure}[t]
    \centering
    
    \definecolor{ink}{HTML}{111827}
    \definecolor{muted}{HTML}{4B5563}
    \definecolor{tealMain}{HTML}{166552}
    \definecolor{tealFill}{HTML}{C7EFE1}
    \definecolor{tealLine}{HTML}{99D5C2}
    \definecolor{rustMain}{HTML}{8C4327}
    \definecolor{rustLine}{HTML}{C6937E}
    \definecolor{softBox}{HTML}{EFECE6}

    \resizebox{0.9\columnwidth}{!}{
        \begin{tikzpicture}[
            >=Stealth,
            every node/.style={text=ink},
            outerbox/.style={
                line width=0.65pt, fill=white, rounded corners=2.5pt,
                minimum width=4.65cm, minimum height=0.72cm,
                font=\scriptsize, align=center, inner sep=2pt
            },
            refsub/.style={
                draw=none, fill=softBox, rounded corners=2pt,
                minimum width=2.08cm, minimum height=0.42cm,
                font=\scriptsize, align=center, inner sep=1pt
            },
            duvsub/.style={
                line width=0.55pt, fill=white, rounded corners=2pt,
                minimum height=0.42cm,
                font=\fontsize{5.8}{6.8}\selectfont, align=center, inner xsep=1.5pt, inner ysep=1pt
            }
        ]

        \node[font=\footnotesize\bfseries] at (2.325, 0)
            {(a) Speculated IR Decomposition};

        \node[outerbox, draw=tealMain] (a_ref) at (2.325, -0.62) {
            \textbf{REF (SoftFloat)}\\[-0.5pt]
            \fontsize{6.5}{7.5}\selectfont\textcolor{muted}{$M$ behavior branches}
        };

        \node[refsub, anchor=west] (a_muladd) at (0.00, -1.32) {\texttt{mulAddF32}};
        \node[font=\scriptsize, text=muted] at (2.325, -1.32) {$\rightarrow$};
        \node[refsub, anchor=east] (a_roundpack) at (4.65, -1.32) {\texttt{roundPackF32}};

        \node[
            draw=tealMain, line width=0.75pt, fill=tealFill, rounded corners=3pt,
            minimum width=3.65cm, minimum height=0.82cm,
            font=\scriptsize, align=center, inner sep=2pt
        ] (a_spec) at (2.325, -2.26) {
            \textbf{Speculated IR $\cdot$ IEEE-754 \texttt{fma}}
        };

        \draw[->, line width=0.6pt, tealMain] (a_muladd.south) -- ($(a_spec.north)+(-0.70, 0)$);
        \draw[->, line width=0.6pt, tealMain] (a_roundpack.south) -- ($(a_spec.north)+(0.70, 0)$);

        \node[duvsub, draw=tealMain, minimum width=1.24cm, anchor=west] (a_m1) at (0.00, -3.20) {MUL};
        \node[duvsub, draw=tealMain, minimum width=1.73cm, anchor=west] (a_m2) at (1.36, -3.20) {FMA-ADD};
        \node[duvsub, draw=tealMain, minimum width=1.44cm, anchor=east] (a_m3) at (4.65, -3.20) {ROUND};

        \draw[->, line width=0.6pt, tealMain] (a_m1.north) -- ($(a_spec.south)+(-0.90, 0)$);
        \draw[->, line width=0.6pt, tealMain] (a_m2.north) -- (a_spec.south);
        \draw[->, line width=0.6pt, tealMain] (a_m3.north) -- ($(a_spec.south)+(0.90, 0)$);

        \node[outerbox, draw=tealMain] (a_duv) at (2.325, -3.90) {
            \textbf{DUV (\texttt{fma\_f32})}\\[-0.5pt]
            \fontsize{6.5}{7.5}\selectfont\textcolor{muted}{$N$ behavior branches}
        };

        \draw[line width=0.45pt, tealLine] (0.05, -4.42) -- (4.60, -4.42);
        \node[anchor=north, font=\scriptsize, align=center, inner sep=0pt] at (2.325, -4.55) {
            \textbf{\textcolor{tealMain}{Proof obligations $\boldsymbol{\approx N + M}$}}\\[1pt]
        };

        \node[font=\footnotesize\bfseries] at (7.325, 0)
            {(b) Structural Decomposition};

        \node[outerbox, draw=rustMain] (b_ref) at (7.325, -0.62) {
            \textbf{REF (SoftFloat)}\\[-0.5pt]
            \fontsize{6.5}{7.5}\selectfont\textcolor{muted}{$M$ behavior branches}
        };

        \node[refsub, anchor=west] (b_muladd) at (5.00, -1.32) {\texttt{mulAddF32}};
        \node[font=\scriptsize, text=muted] at (7.325, -1.32) {$\rightarrow$};
        \node[refsub, anchor=east] (b_roundpack) at (9.65, -1.32) {\texttt{roundPackF32}};

        \node[duvsub, draw=rustMain, densely dashed, minimum width=1.24cm, anchor=west] (b_m1) at (5.00, -3.20) {MUL};
        \node[duvsub, draw=rustMain, densely dashed, minimum width=1.73cm, anchor=west] (b_m2) at (6.36, -3.20) {FMA-ADD};
        \node[duvsub, draw=rustMain, densely dashed, minimum width=1.44cm, anchor=east] (b_m3) at (9.65, -3.20) {ROUND};

        \foreach \target in {b_m1.north, b_m2.north, b_m3.north} {
            \draw[line width=0.6pt, rustMain] (b_muladd.south) -- (\target);
            \draw[line width=0.6pt, rustMain] (b_roundpack.south) -- (\target);
        }

        \node[outerbox, draw=rustMain] (b_duv) at (7.325, -3.90) {
            \textbf{DUV (\texttt{fma\_f32})}\\[-0.5pt]
            \fontsize{6.5}{7.5}\selectfont\textcolor{muted}{$N$ behavior branches}
        };

        \draw[line width=0.45pt, rustLine] (5.05, -4.42) -- (9.60, -4.42);
        \node[anchor=north, font=\scriptsize, align=center, inner sep=0pt] at (7.325, -4.55) {
            \textbf{\textcolor{rustMain}{Proof obligations $\boldsymbol{\approx N \times M}$}}\\[1pt]
        };

        \end{tikzpicture}
    }
    \vspace{-10pt}
    \caption{Comparison of proof decomposition strategies for \texttt{fma\_f32}. (a) Speculated IR decomposition. (b) Structural decomposition.}
    \label{fig:fma_decomposition}
\end{figure}

%% file: sections/2_methods.tex
\input{figs/fig-chain}

\section{Method}
\label{sec:method}

TheoremV verifies a design under verification (DUV, e.g., RTL Verilog/SystemVerilog) against a reference model (REF, e.g., a C implementation or formal ISA specification), optionally assisted by specification documents (Fig.~\ref{fig:framework_overview}). It compiles both into executable Lean~4 models, inserts \textit{Speculated IRs} to decompose verification along semantic abstraction levels, and dispatches adjacent proofs to subagents. A main agent coordinates this chain and composes the proofs into an end-to-end equivalence theorem or reports a confirmed counterexample.

To bridge the abstraction gap between the DUV and REF (gap~\ding{182}), Speculated IRs provide intermediate models (Section~\ref{sec:method:ir}) linking circuit behavior to underlying algorithms and specifications, thereby reducing the semantic distance covered by each proof. Our agentic workflow (Section~\ref{sec:method:workflow}) automatically \textit{proposes, filters, and revises} these initially unverified IRs and dispatches the resulting subgoals to subagents in parallel.

To bridge the representation gap between RTL and Lean (gap~\ding{183}), our RTL-to-Lean compiler (Section~\ref{sec:method:translation}) ensures broad Verilog semantic coverage, maps parallel circuit behavior into dependency-ordered pure functions, and preserves module definitions for reuse across instances. With the translators that convert high-level reference models into Lean, including a C-to-Lean and Sail-to-Lean translator, DUV and REF are in the same proof environment.
\subsection{Speculated IR}
\label{sec:method:ir}

A Speculated IR is an executable intermediate model proposed by the agent to bridge semantic gaps between adjacent levels of abstraction.
By inserting intermediate models between the DUV and REF, the framework decomposes a monolithic proof spanning wide semantic gaps into a chain of local equivalence or refinement proofs.
Depending on the verification bottleneck, a Speculated IR may capture either the entire design or an isolated subfunction, with its scope and abstraction level guided by the behavior of the DUV and REF.

\textbf{An illustrative example of semantic decomposition.}
Fig.~\ref{fig:gap} illustrates the semantic gap between two implementations of floating-point division. In Fig.~\ref{fig:gap}(a), the REF divides the scaled significand \texttt{mA64} by \texttt{mB} in a single integer operation before rounding. In Fig.~\ref{fig:gap}(b), the DUV computes the quotient iteratively over multiple cycles in an SRT core, selecting quotient digits step by step while maintaining the partial remainder in carry-save form. Because the REF exposes no intermediate states matching these cycle-by-cycle updates, direct state-level equivalence checking fails.

\input{figs/fig-gap}

To bridge this gap, we introduce two Speculated IRs in Fig.~\ref{fig:gap}(c): an algorithmic IR ($\mathrm{IR}_{\mathrm{SRT}}$) that replaces the multi-cycle SRT core and carry-save representation with an integer recurrence, and a specification IR ($\mathrm{IR}_{754}$) that formalizes exact IEEE~754 division and rounding semantics. For the normal-number path under the default RNE rounding mode shown, letting $\equiv$ denote bit-exact output equivalence on identical inputs, we establish the proof chain
\(
  \mathrm{DUV} \equiv \mathrm{IR}_{\mathrm{SRT}} \equiv \mathrm{IR}_{754} \equiv \mathrm{REF},
\)
where each intermediate model targets a distinct abstraction barrier:

(1)~\emph{State abstraction via $\mathrm{IR}_{\mathrm{SRT}}$.} While structural decomposition isolates the SRT core, proving its functional correctness requires lifting its low-level hardware encodings into mathematical values. Specifically, the DUV maintains two redundant quotient candidates (\texttt{Q}, \texttt{Q\_m1}) alongside a carry-save remainder (\texttt{ws}, \texttt{wc}), whereas $\mathrm{IR}_{\mathrm{SRT}}$ tracks the abstract partial remainder $W$ and partial quotient $q$ as integers (e.g., updating $q$ to $16q + 4d_1 + d_2$ across a full iteration with two radix-4 digits $d_1, d_2$). The local proof $\mathrm{DUV} \equiv \mathrm{IR}_{\mathrm{SRT}}$ verifies that decoding the initial registers recovers $(W, q)$ and that each \texttt{ITER} cycle preserves this relation under the recurrence update.

(2)~\emph{Specification abstraction via $\mathrm{IR}_{754}$.} Having reduced the multi-cycle SRT core to an integer recurrence, we must still verify that the surrounding divider datapath and rounding logic match the REF. Rather than proving equivalence directly between two distinct algorithms, $\mathrm{IR}_{754}$ acts as a neutral specification anchor defining the target IEEE~754 quotient for any operand pair and rounding mode. This requires two separate proofs: showing that significand division and rounding in $\mathrm{IR}_{\mathrm{SRT}}$ satisfy $\mathrm{IR}_{754}$, and proving that the REF's single-step integer division and rounding satisfy the same specification.

Composing these three local proofs by transitivity guarantees that the multi-cycle DUV and the REF produce identical output bits for every input on the illustrated path.

\textbf{Roles of Speculated IRs.}
Generalizing from this example, Speculated IRs bridge semantic gaps across diverse hardware designs through three complementary forms of abstraction:
\begin{itemize}
    \item \textbf{State Abstraction.} Hardware implementations often rely on low-level encodings (e.g., redundant number systems or one-hot states) that obscure arithmetic or logical intent. An IR can lift bit-level register states to abstract mathematical objects while preserving the original control structure, as $\mathrm{IR}_{\mathrm{SRT}}$ does by mapping encoded registers to $(W, q)$.
    \item \textbf{Temporal Abstraction.} Microarchitectural optimizations frequently spread a single architectural operation across multiple pipeline stages or clock cycles. An IR can compress multi-cycle execution into coarser atomic steps to align observation boundaries with the REF. For instance, in the PicoRV32 RISC-V processor, a single-cycle IR abstracts multi-cycle instruction execution into one step per instruction, enabling direct state comparison against the Sail ISA model at instruction boundaries.
    \item \textbf{Specification Abstraction.} When the DUV and REF implement the same high-level behavior through vastly different algorithms, a specification IR serves as a neutral, declarative anchor. Rather than proving direct equivalence between two complex implementations, both sides are proved to satisfy this common specification, as seen with $\mathrm{IR}_{754}$ in the divider.
\end{itemize}
Guided by the DUV, REF, and optional specification documents, the main agent determines which abstractions are needed and orders them into a progressive refinement chain.

\subsection{Parallel Agentic Proving with Speculated IRs}
\label{sec:method:workflow}

Because LLM-generated Speculated IRs are heuristic proposals rather than guaranteed abstractions, an initial candidate may contain subtle semantic bugs or leave an abstraction gap too wide to prove in one step.
Attempting formal proof in Lean directly on flawed models is computationally expensive and wastes substantial proving budget.
To address this, we make every Speculated IR \emph{executable}, pairing fast differential testing with parallel formal proving in an iterative \textit{propose--filter--revise} workflow.

\textbf{Iterative chain construction by the main agent.}
The main agent orchestrates the overall proof structure through an iterative \textit{propose--filter--revise} loop:
(1)~\textbf{Propose.} Inspecting the DUV, REF, and any available specification documents, the main agent identifies underlying algorithms, data encodings, and control boundaries, then synthesizes an executable Lean model for each Speculated IR along with local equivalence or refinement theorem statements between adjacent models.
(2)~\textbf{Filter.} Before invoking expensive proof search, the main agent runs lightweight differential testing by executing candidate IRs alongside adjacent Lean models on randomized inputs and comparing outputs at matching observation boundaries (checking local subfunction IRs against the corresponding DUV module, or embedding them in a surrounding wrapper for full-design checks). This rapid check catches observable semantic errors within seconds, ensuring that costly formal proving is spent only on empirically consistent candidates.
(3)~\textbf{Revise.} The main agent updates the chain using feedback from either differential filtering or subagent proving. When a testing mismatch or proof counterexample exposes an inaccurate IR definition or interface, the agent repairs the model; when a subagent proof times out due to an overly large semantic leap, the agent subdivides that edge by inserting a finer-grained intermediate IR. Any revised model must pass differential filtering again before proof tasks are dispatched.

\textbf{Parallel proving and counterexample triage by subagents.}
Once adjacent interfaces pass filtering, the main agent dispatches each adjacent model pair to an isolated subagent.
Because each edge spans only a single abstraction step, subagents operate within a compact proof context and execute in parallel, substantially reducing end-to-end proving latency.
Each subagent returns a kernel-checked Lean proof, a proof bottleneck report (e.g., timeout), or a concrete counterexample.
When a counterexample is returned, the main agent validates whether the discrepancy is reproducible on a valid top-level input trace:
(1)~\textbf{IR or interface error:} if the DUV and REF agree while the IR diverges—or if the local counterexample is unreachable from valid top-level states—the main agent triggers a revision of the IR definition or its interface assumptions and resumes proving for the affected edges;
(2)~\textbf{Confirmed design bug:} if the discrepancy reproduces end-to-end between the DUV and REF, the framework terminates early and reports a confirmed DUV--REF mismatch.

\textbf{End-to-end proof composition.}
Once all adjacent edges in the chain are proved, the main agent composes the local proofs by transitivity into the final end-to-end theorem and checks the complete proof with the Lean kernel, guaranteeing formal equivalence or refinement between the DUV and REF.

\subsection{Hierarchy-Preserving Compilation to Lean Models}
\label{sec:method:translation}

To bridge the representation gap between hardware designs and Lean (gap~\ding{183}), TheoremV compiles both the DUV and REF into executable, hierarchy-preserving Lean~4 state machines in one proof environment. Pure functional execution enables fast differential filtering before proving, while preserving module hierarchy prevents code bloat and lets agents isolate subfunction Speculated IRs.

\textbf{Word-level lowering to executable state machines.}
Built on Yosys~\cite{yosys}, our RTL-to-Lean compiler elaborates Verilog/SystemVerilog into Yosys's word-level intermediate representation, \textbf{RTLIL}, reusing mature EDA handling of macros, \texttt{generate} blocks, and parameters while mapping roughly 40 cell types directly to Lean~4's fixed-width bit-vector (\texttt{BitVec}) operations without premature bit-blasting. To model concurrent hardware in a sequential proof language~\cite{loow2022}, the compiler translates each module into a deterministic, purely functional state machine: combinational cells are topologically ordered into an \texttt{outputs} function, while state updates are isolated into a cycle-accurate \texttt{step} function. 
This decoupling exposes clean inductive structure for proving while enabling fast native Lean execution for differential filtering.

\textbf{Hierarchy preservation via port-level dependency splitting.}
Flattening multi-module designs causes severe code bloat and obscures modular boundaries, so we preserve module hierarchy as reusable per-module Lean definitions. However, invoking a submodule as one monolithic function creates false combinational loops when a parent feeds one output back to an independent input, even if internal logic is acyclic. We eliminate these false loops via per-output dependency splitting (Figure~\ref{fig:falseloop}): the compiler identifies which inputs or registers affect each output port and evaluates each output separately in the parent, passing only relevant inputs and zeroing unused ones. This breaks false loops at module boundaries while preserving hierarchy and executable semantics.

\input{figs/fig-falseloop}

\textbf{Reference model translation.}
For end-to-end proofs in one environment, TheoremV pairs the RTL compiler with two REF translation paths: a lightweight C/C++-to-Lean translator lowering algorithmic reference models (such as Berkeley SoftFloat~\cite{softfloat} and NVDLA C++ models~\cite{nvdla}) into matching pure Lean functions, and formal ISA backends such as Sail~\cite{sail} exporting RISC-V semantics directly to Lean~4. We thoroughly validate both our RTL and C/C++ translators against standard simulators and native executables to ensure faithful translation.

%% file: figs/fig-chain.tex
\definecolor{chainFail}{RGB}{188,48,48}
\definecolor{chainPass}{RGB}{28,130,72}
\definecolor{leanBlue}{RGB}{32,68,128}
\definecolor{irAmber}{RGB}{165,95,20}

\begin{figure*}[!htbp]
\centering
\resizebox{\linewidth}{!}{
\begin{tikzpicture}[
  >=Latex,
  card/.style={
    draw=black!70, fill=white, rounded corners=3.5pt, align=center,
    inner sep=3pt, line width=0.8pt
  },
  chainnode/.style={minimum height=9.6mm, text width=2.22cm},
  compnode/.style={minimum height=9.6mm, text width=2.45cm},
  workflowstep/.style={minimum height=9.5mm},
  proofwork/.style={minimum height=15.5mm, inner sep=3.5pt},
  dut/.style={card, fill=white, draw=leanBlue!75},
  rtlil/.style={card, fill=white, draw=leanBlue!75},
  ref/.style={card, fill=white, draw=leanBlue!75},
  dutlean/.style={card, fill=leanBlue!6!white, draw=leanBlue!90},
  reflean/.style={card, fill=leanBlue!6!white, draw=leanBlue!90},
  ir/.style={card, fill=irAmber!7!white, draw=irAmber!85},
  agent/.style={
    card, fill=leanBlue!13!white, draw=leanBlue, line width=1.05pt
  },
  subbox/.style={
    draw=leanBlue!65, fill=white, rounded corners=3pt,
    align=center, inner sep=3.5pt, line width=0.75pt
  },
  passrow/.style={
    draw=leanBlue!28, fill=leanBlue!2.5!white, rounded corners=2pt,
    minimum width=5.56cm, minimum height=4.8mm, inner sep=0pt
  },
  decision/.style={
    diamond, aspect=1.85, draw=irAmber!90, fill=irAmber!6!white,
    align=center, inner sep=0.6pt, line width=0.85pt, font=\footnotesize\bfseries
  },
  stepnum/.style={
    circle, fill=leanBlue, text=white, inner sep=1.5pt,
    font=\small\bfseries
  },
  subbadge/.style={
    fill=leanBlue!10!white, draw=leanBlue!55, rounded corners=2pt,
    inner xsep=5pt, inner ysep=1.2pt, font=\footnotesize\bfseries, text=leanBlue
  },
  trans/.style={
    ->, draw=black!75, line width=0.9pt,
    >={Latex[length=2.0mm, width=1.35mm]}
  },
  proof/.style={
    <->, draw=leanBlue!90, line width=0.95pt,
    >={Latex[length=1.85mm, width=1.25mm]}
  },
  ce_arrow/.style={
    ->, draw=chainFail, densely dashed, line width=0.95pt, rounded corners=3.5pt,
    >={Latex[length=2.0mm, width=1.35mm]}
  },
  pass_arrow/.style={
    ->, draw=chainPass, line width=0.95pt, rounded corners=3.5pt,
    >={Latex[length=2.0mm, width=1.35mm]}
  },
  note/.style={font=\footnotesize, text=black!85, align=center},
  pnote/.style={
    font=\small\bfseries, text=leanBlue, fill=white,
    inner sep=1.2pt, rounded corners=1.5pt
  },
  pill/.style={
    fill=white, inner xsep=3.5pt, inner ysep=1pt, rounded corners=2pt
  }
]

\node[font=\normalsize\bfseries, text=black!90] (b1_title) at (4.59, -0.36)
  {\tikz[baseline=-0.6ex]\node[stepnum]{1}; RTL/REF-to-Lean Compilation};

\node[dut, compnode] (b1_vdut) at (3.16, -1.14)
  {\textbf{\small RTL DUV}\\[1pt]\footnotesize Verilog / SV};

\node[ref, compnode] (b1_gref) at (7.65, -1.14)
  {\textbf{\small Golden REF}\\[1pt]\footnotesize C (SoftFloat) / Sail};

\node[rtlil, compnode] (b1_rtlil) at (3.16, -2.46)
  {\textbf{\small RTLIL}\\[1pt]\footnotesize Word-level IR};

\node[rtlil, compnode] (b1_reftrans) at (7.65, -2.46)
  {\textbf{\small C / Sail}\\[1pt]\footnotesize Translators};

\draw[trans] (b1_vdut.south) --
  node[right=3pt, note] {Elaborate} (b1_rtlil.north);

\draw[trans] (b1_gref.south) --
  node[right=3pt, note] {} (b1_reftrans.north);

\node[subbox, minimum width=5.82cm, minimum height=1.84cm, inner sep=0pt]
  (b1_methods) at (3.16, -4.14) {};

\foreach \i/\ltxt/\rtxt in {
  1/{Word-level cells}/{BitVec operations},
  2/{Sequential logic}/{Pure-functional FSM},
  3/{Module hierarchy}/{Dependency splitting}
} {
  \pgfmathsetmacro{\ypos}{-2.99 - 0.575*\i}
  \node[passrow] at (3.16, \ypos) {};
  \node[note, text height=1.5ex, text depth=0.35ex, anchor=west] at (0.55, \ypos) {\ltxt};
  \draw[trans] (2.74, \ypos) -- (3.04, \ypos);
  \node[note, text=leanBlue, text height=1.5ex, text depth=0.35ex, anchor=west] at (3.18, \ypos) {\rtxt};
}

\draw[trans] (b1_rtlil.south) -- (b1_methods.north);

\node[dutlean, compnode] (b1_ldut) at (3.16, -5.82)
  {\textbf{\small DUV in Lean}\\[1pt]\footnotesize $\mathcal{M}_{\text{DUV}}$};

\node[reflean, compnode] (b1_lref) at (7.65, -5.82)
  {\textbf{\small REF in Lean}\\[1pt]\footnotesize $\mathcal{M}_{\text{REF}}$};

\draw[trans] (b1_methods.south) -- (b1_ldut.north);
\draw[trans] (b1_reftrans.south) -- (b1_lref.north);

\begin{scope}[on background layer]
  \draw[draw=black!50, fill=black!1.5, rounded corners=5pt, line width=0.85pt]
    (0, 0) rectangle (9.18, -6.95);
\end{scope}

\begin{scope}[shift={(9.53, 0)}]
  \node[font=\normalsize\bfseries, text=leanBlue] (b2_title) at (7.50, -0.35)
    {\tikz[baseline=-0.6ex]\node[stepnum]{2}; Counterexample-Guided Agentic Verification in Lean 4};

  \node[dutlean, chainnode] (c_duv) at (1.75, -1.28) {
    \textbf{\small DUV in Lean}\\[1pt]
    \footnotesize $\mathcal{M}_{\text{DUV}}$
  };

  \node[ir, chainnode] (c_ir1) at (5.45, -1.28) {
    \textbf{\small $\mathbf{IR_1}$ model}\\[1pt]
    \footnotesize Algorithmic
  };

  \node[ir, chainnode] (c_ir2) at (9.15, -1.28) {
    \textbf{\small $\mathbf{IR_2}$ model}\\[1pt]
    \footnotesize Specification
  };

  \node[font=\small\bfseries, text=leanBlue, inner sep=1pt] (c_dots) at (10.90, -1.28) {$\cdots$};

  \node[reflean, chainnode] (c_ref) at (13.25, -1.28) {
    \textbf{\small REF in Lean}\\[1pt]
    \footnotesize $\mathcal{M}_{\text{REF}}$
  };

  \draw[proof] (c_duv) -- node[pnote, above=1.2pt] {$\sim_1$}
                          node[note, below=1.2pt] {$\pi_1$} (c_ir1);
  \draw[proof] (c_ir1) -- node[pnote, above=1.2pt] {$\sim_2$}
                          node[note, below=1.2pt] {$\pi_2$} (c_ir2);
  \draw[draw=leanBlue!90, line width=0.95pt] (c_ir2) -- (c_dots);
  \draw[proof] (c_dots) -- node[pnote, above=1.2pt] {$\sim_k$}
                           node[note, below=1.2pt] {$\pi_k$} (c_ref);

  \node[agent, workflowstep, text width=1.60cm] (main_agent) at (1.55, -4.22) {
    \textbf{\small Main}\\[-1pt]\textbf{\small Agent}
  };

  \node[ir, workflowstep, text width=1.85cm] (spec_ir) at (4.35, -4.22) {
    \textbf{\small Speculated}\\[-1pt]\textbf{\small IR model}
  };
  \draw[trans] (main_agent.east) -- (spec_ir.west);

  \node[decision, minimum width=1.95cm, minimum height=1.10cm] (difftest) at (7.05, -4.22) {
    Difftest\\[-1pt]Success?
  };
  \draw[trans] (spec_ir.east) -- (difftest.west);

  \draw[ce_arrow] (difftest.north) -- (7.05, -2.42)
    -- node[pill, font=\small\bfseries, text=chainFail] {Fail (Revise IR)}
    (1.55, -2.42) -- (main_agent.north);

  \node[subbox, proofwork, text width=1.95cm] (sub1) at (10.65, -3.16) {
    \vspace*{4pt}
    \textbf{\small DUV}\\[-1pt]
    $\Updownarrow$\\[-1pt]
    \textbf{\small IR model}
  };
  \node[subbadge] at (sub1.north) {SubAgent$_1$};

  \node[subbox, proofwork, text width=1.95cm] (sub2) at (10.65, -5.28) {
    \vspace*{4pt}
    \textbf{\small IR model}\\[-1pt]
    $\Updownarrow$\\[-1pt]
    \textbf{\small REF}
  };
  \node[subbadge] at (sub2.north) {SubAgent$_2$};

  \coordinate (dispatch_hub) at (8.90, -4.22);
  \draw[pass_arrow] (difftest.east) --
    node[above=1.5pt, font=\small\bfseries, text=chainPass] {$\surd$}
    (dispatch_hub);
  \draw[pass_arrow] (dispatch_hub) |- (sub1.west);
  \draw[pass_arrow] (dispatch_hub) |- (sub2.west);

  \draw[ce_arrow] (sub2.south) -- (10.65, -6.36)
    -- node[pill, font=\small\bfseries, text=chainFail] {Counterexample / Timeout (Revise IR)}
    (1.55, -6.36) -- (main_agent.south);

  \node[subbox, minimum height=19.5mm, inner sep=4pt, fill=chainPass!5, draw=chainPass, line width=0.85pt, text width=1.88cm]
    (assembled) at (13.55, -4.22) {
    \textbf{\small DUV}\\[-1pt]
    $\Updownarrow$\\[-1pt]
    \textbf{\small IR model}\\[-1pt]
    $\Updownarrow$\\[-1pt]
    \textbf{\small REF}
  };

  \coordinate (asm_hub) at (12.06, -4.22);
  \draw[draw=chainPass, line width=0.95pt, rounded corners=3pt]
    (sub1.east) -| (asm_hub) |- (sub2.east);
  \draw[trans, draw=chainPass] (asm_hub) -- (assembled.west);

  \begin{scope}[on background layer]
    \draw[
      fill=leanBlue!3!white, draw=leanBlue!65, rounded corners=5pt,
      line width=0.85pt, dashed
    ] (0, 0) rectangle (15.00, -6.95);
    \draw[
      fill=white, draw=leanBlue!45, rounded corners=4pt,
      line width=0.75pt
    ] (0.35, -0.66) rectangle (14.65, -1.90);
  \end{scope}
\end{scope}

\coordinate (handoff_bus) at (9.355, -4.22);
\draw[draw=leanBlue!85, line width=0.9pt]
  (b1_ldut.south) -- (3.16, -6.64) -- (9.355, -6.64) -- (handoff_bus)
  (b1_lref.south) -- (7.65, -6.64);
\draw[->, draw=leanBlue!85, line width=0.95pt, >={Latex[length=2.0mm, width=1.35mm]}]
  (handoff_bus) -- (main_agent.west);

\end{tikzpicture}
}
\caption{Overview of TheoremV. \textbf{(1) Front-end translation:} Verilog/SystemVerilog DUVs and C/Sail reference models are translated into executable Lean~4 state machines. \textbf{(2) Agentic proof decomposition:} A counterexample-guided workflow proposes, differentially filters, and revises \textit{Speculated IRs} to decompose the equivalence proof into local refinement lemmas, which subagents prove in parallel for the main agent to compose the whole proof.}
\label{fig:framework_overview}
\end{figure*}
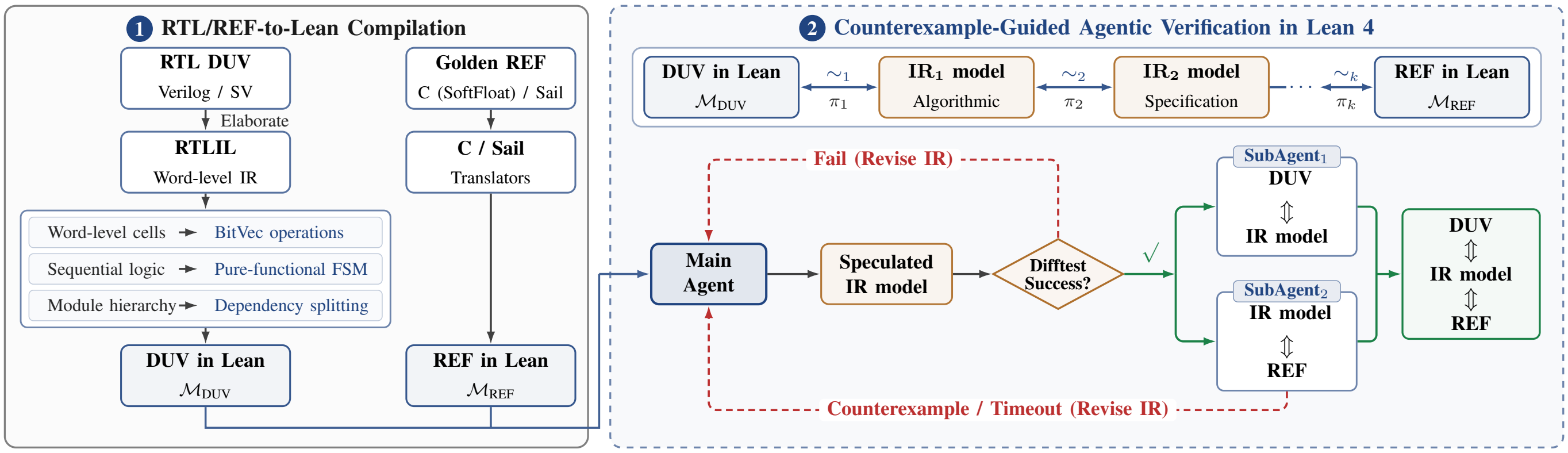

%% file: figs/fig-gap.tex
\definecolor{kwblue}{RGB}{0,0,180}
\definecolor{funcviolet}{RGB}{120,0,175}
\definecolor{commentgreen}{RGB}{0,105,0}
\definecolor{numgray}{gray}{0.45}
\definecolor{bordergray}{gray}{0.45}
\definecolor{irSRTbg}{RGB}{243,235,255}
\definecolor{irSRTbd}{RGB}{120,40,200}
\definecolor{irIEEEbg}{RGB}{232,246,255}
\definecolor{irIEEEbd}{RGB}{0,110,190}
\definecolor{statebg}{RGB}{235,240,248}

\newcommand{\stbadge}[1]{
  \colorbox{statebg}{\textcolor{kwblue}{\textbf{\texttt{#1}}}}}
\lstdefinestyle{cleanAlgo}{
  basicstyle=\fontsize{5.8}{7.6}\selectfont\ttfamily,
  numbers=left,
  numberstyle=\fontsize{4.8}{7.6}\selectfont\ttfamily\color{numgray},
  numbersep=3pt,xleftmargin=8pt,aboveskip=1pt,belowskip=0pt,
  showstringspaces=false,columns=flexible,keepspaces=true,
  escapeinside={(*@}{@*)},keywordstyle=\color{kwblue}\bfseries,
  morekeywords={if,then,else,return,init},
  keywordstyle=[2]\color{funcviolet}\bfseries,
  morekeywords=[2]{unpack,pack,DigitSelect,CSA_Step,OTFC_Step,signed},
  commentstyle=\color{commentgreen}\itshape,morecomment=[l]{//}}
\tcbset{algocard/.style={
  enhanced,colback=white,colframe=bordergray,boxrule=0.55pt,arc=3pt,
  left=2.5pt,right=2.5pt,top=2.5pt,bottom=2.5pt,boxsep=1pt,
  fonttitle=\fontsize{6.6}{8}\selectfont\sffamily\bfseries,
  coltitle=black,colbacktitle=black!4,titlerule=0.35pt,
  titlerule style={bordergray},before skip=0pt,after skip=0pt}}

\newcommand{\fdivmark}[1]{\tikz[remember picture,baseline=(#1.base)]\coordinate (#1);}

\newsavebox{\fdivLeftBox}
\newsavebox{\fdivDutBox}
\newsavebox{\fdivFullBox}
\newlength{\fdivLeftHeight}
\newlength{\fdivRightHeight}
\newlength{\fdivPanelHeight}

\begin{figure*}[!tbp]
\centering

\begin{lrbox}{\fdivLeftBox}
\begin{minipage}[t]{0.345\textwidth}
\begin{tcolorbox}[algocard,title={(a) REF: sequential C}]
\fontsize{5.8}{7.4}\selectfont\sffamily
\textbf{In:} \texttt{a,b : float32}\hfill\textbf{Out:} \texttt{z : float32}
\vspace{2pt}\hrule height 0.35pt\vspace{2pt}
\begin{lstlisting}[style=cleanAlgo]
(sZ,eZ,mA,mB) (*@$\leftarrow$@*) unpack(a,b)
if mA < mB then
  eZ (*@$\leftarrow$@*) eZ - 1; mA64 (*@$\leftarrow$@*) uint64(mA)<<31
else mA64 (*@$\leftarrow$@*) uint64(mA)<<30
(*@\fdivmark{L_top}@*)mZ (*@$\leftarrow$@*) mA64 (*@$\div$@*) mB
if (mZ & 0x3F) == 0 then
  mZ (*@$\leftarrow$@*) mZ | (mB (*@$\times$@*) mZ (*@$\neq$@*) mA64)
roundBits (*@$\leftarrow$@*) mZ & 0x7F
mZ (*@$\leftarrow$@*) (mZ + 0x40) (*@$\gg$@*) 7
if roundBits == 0x40 then mZ (*@$\leftarrow$@*) mZ & ~1(*@\fdivmark{L_bot}@*)
return pack(sZ,eZ,mZ)
\end{lstlisting}
\end{tcolorbox}
\vskip 4.5pt
\begin{tcolorbox}[algocard,colback=black!2.5,title={(d) Proved full-design equivalence}]
\fontsize{5.8}{7.2}\selectfont\sffamily\centering
$\forall\,\textsf{32-bit } a,b,\mathit{rm}:\; \mathrm{DUV}(a,b,\mathit{rm}) \equiv \mathrm{REF}(a,b,\mathit{rm})$
\end{tcolorbox}
\end{minipage}
\end{lrbox}

\begin{lrbox}{\fdivDutBox}
\begin{minipage}[t]{0.415\textwidth}
\begin{tcolorbox}[algocard,title={(b) DUV: synchronous RTL}]
\fontsize{5.6}{7.4}\selectfont\sffamily
\textbf{In:} \texttt{[31:0] a,b; in\_valid}
\hfill \textbf{Out:} \texttt{[31:0] result}
\vspace{2pt}\hrule height 0.35pt\vspace{2pt}
\begin{lstlisting}[style=cleanAlgo]
 (*@\stbadge{IDLE}@*): if in_valid: (sZ,eZ,mA,mB) (*@$\leftarrow$@*) unpack(a,b)
 (*@\stbadge{PRE\_1}@*): ws<=(mA<mB)?mA<<2:mA<<1; wc<=0; state<=ITER
  init qPrev,Q,Q_m1; cnt <= 6
 (*@\stbadge{ITER}@*) // 6 full cycles (2 digits) + 1 half
  (*@\fdivmark{Rsrt_top}@*)(q1,q2) (*@$\leftarrow$@*) DigitSelect(ws,wc,qPrev,mB)
  (ws,wc) <= CSA_Step(ws,wc,mB,qPrev,q1,q2,cnt==0)(*@\fdivmark{Rsrt_maxr}@*)
  (Q,Q_m1) <= OTFC_Step(Q,Q_m1,qPrev,q1,q2,cnt==0)
  qPrev<=q2; cnt<=cnt-1; if !cnt: state<=POST_0(*@\fdivmark{Rsrt_bot}@*)
 (*@\stbadge{POST\_0}@*)
  (*@\fdivmark{Rieee_top}@*)rem (*@$\leftarrow$@*) signed(ws+wc); qFinal (*@$\leftarrow$@*) (rem<0)?Q_m1:Q(*@\fdivmark{Rieee_maxr}@*)
  rnd (*@$\leftarrow$@*) qFinal[0]; stk (*@$\leftarrow$@*) (rem != 0)
  up (*@$\leftarrow$@*) rnd & (stk | qFinal[1])
  (cout,mant) (*@$\leftarrow$@*) qFinal[23:1] + up(*@\fdivmark{Rieee_bot}@*)
  result <= pack(sZ,eZ+cout,mant); state<=IDLE
\end{lstlisting}
\end{tcolorbox}
\end{minipage}
\end{lrbox}

\setlength{\fdivLeftHeight}{\dimexpr\ht\fdivLeftBox+\dp\fdivLeftBox\relax}
\setlength{\fdivRightHeight}{\dimexpr\ht\fdivDutBox+\dp\fdivDutBox\relax}
\ifdim\fdivLeftHeight>\fdivRightHeight
  \setlength{\fdivPanelHeight}{\fdivLeftHeight}
\else
  \setlength{\fdivPanelHeight}{\fdivRightHeight}
\fi

\begin{lrbox}{\fdivFullBox}
\begin{minipage}{\textwidth}
\begin{tabular}{@{}p{0.345\textwidth}@{}p{0.24\textwidth}@{}p{0.415\textwidth}@{}}
\begin{minipage}[t][\fdivPanelHeight][s]{\linewidth}
\vspace{0pt}
\begin{tcolorbox}[algocard,title={(a) REF: sequential C}]
\fontsize{5.8}{7.4}\selectfont\sffamily
\textbf{In:} \texttt{a,b : float32}\hfill\textbf{Out:} \texttt{z : float32}
\vspace{2pt}\hrule height 0.35pt\vspace{2pt}
\begin{lstlisting}[style=cleanAlgo]
(sZ,eZ,mA,mB) (*@$\leftarrow$@*) unpack(a,b)
if mA < mB then
  eZ (*@$\leftarrow$@*) eZ - 1; mA64 (*@$\leftarrow$@*) uint64(mA)<<31
else mA64 (*@$\leftarrow$@*) uint64(mA)<<30
(*@\fdivmark{L_top}@*)mZ (*@$\leftarrow$@*) mA64 (*@$\div$@*) mB
if (mZ & 0x3F) == 0 then
  mZ (*@$\leftarrow$@*) mZ | (mB (*@$\times$@*) mZ (*@$\neq$@*) mA64)
roundBits (*@$\leftarrow$@*) mZ & 0x7F
mZ (*@$\leftarrow$@*) (mZ + 0x40) (*@$\gg$@*) 7
if roundBits == 0x40 then mZ (*@$\leftarrow$@*) mZ & ~1(*@\fdivmark{L_bot}@*)
return pack(sZ,eZ,mZ)
\end{lstlisting}
\end{tcolorbox}
\vfill
\begin{tcolorbox}[algocard,colback=black!2.5,title={(d) Proved full-design equivalence},remember as=boxD]
\fontsize{5.8}{7.2}\selectfont\sffamily\centering
$\forall\,\textsf{32-bit } a,b,\mathit{rm}:\; \mathrm{DUV}(a,b,\mathit{rm}) \equiv \mathrm{REF}(a,b,\mathit{rm})$
\end{tcolorbox}
\end{minipage}
&
\begin{minipage}[t]{\linewidth}
\vspace{0pt}
\hspace*{5pt}\begin{minipage}[t][\fdivPanelHeight][s]{\dimexpr\linewidth-10pt\relax}
\vspace{0pt}
\noindent\tikz[overlay]{
  \fill[black!4,rounded corners=3pt] (0,0) rectangle (\linewidth,-10pt);
  \draw[bordergray,line width=0.55pt,rounded corners=3pt]
    (0,0) rectangle (\linewidth,-\fdivPanelHeight);
  \draw[bordergray,line width=0.35pt] (0,-10pt) -- (\linewidth,-10pt);
}\par
\vspace{-1.5pt}
\noindent\hspace*{3pt}\fontsize{6.6}{8}\selectfont\sffamily\bfseries (c) Speculated IRs\par
\vspace{4pt}
\vfill
\centering
\begin{tikzpicture}[remember picture,
  ircard/.style={rounded corners=2.5pt,line width=0.55pt,
    text width=0.90\linewidth,align=center,inner sep=3.5pt}]
\node[ircard,draw=irSRTbd,fill=irSRTbg] (ISRT) {
  \fontsize{6.2}{7.5}\selectfont\sffamily
  \textbf{$\mathrm{IR}_{\mathrm{SRT}}$ (Algorithmic Recurrence)}\\[3pt]
  \fontsize{5.7}{7.5}\selectfont
  \textbf{Lift:} \texttt{(ws,wc),(Q,Q\_m1)} $\mapsto (W,q)$\\[2pt]
  \textbf{Step:} $q \leftarrow 16q + 4q_1 + q_2$\\[2pt]
  \textbf{Contract:} integer $(W,q)$ quotient/remainder
};
\node[ircard,draw=irIEEEbd,fill=irIEEEbg,below=10pt of ISRT] (I754) {
  \fontsize{6.2}{7.5}\selectfont\sffamily
  \textbf{$\mathrm{IR}_{754}$ (Exact Div.\ \& Rounding)}\\[3pt]
  \fontsize{5.7}{7.5}\selectfont
  \textbf{Exact:} rational quotient $m_A / m_B$\\[2pt]
  \textbf{Align:} scaled significand shifts\\[2pt]
  \textbf{Round:} canonical $\mathrm{RNE}(m_A / m_B)$
};
\draw[bordergray,->,>=Stealth,line width=0.55pt] (ISRT.south) --
  node[midway,fill=white,inner sep=1.2pt,
       font=\fontsize{5.4}{6.2}\selectfont\sffamily\bfseries\itshape,
       text=black!75] {exports contract}
  (I754.north);
\end{tikzpicture}\par
\vfill
\vspace{14pt}\mbox{}
\end{minipage}
\end{minipage}
&
\begin{minipage}[t]{\linewidth}
\vspace{0pt}
\begin{tcolorbox}[algocard,title={(b) DUV: synchronous RTL},height=\fdivPanelHeight]
\fontsize{5.6}{7.4}\selectfont\sffamily
\textbf{In:} \texttt{[31:0] a,b; in\_valid}
\hfill \textbf{Out:} \texttt{[31:0] result}
\vspace{2pt}\hrule height 0.35pt\vspace{2pt}
\begin{lstlisting}[style=cleanAlgo]
 (*@\stbadge{IDLE}@*): if in_valid: (sZ,eZ,mA,mB) (*@$\leftarrow$@*) unpack(a,b)
 (*@\stbadge{PRE\_1}@*): ws<=(mA<mB)?mA<<2:mA<<1; wc<=0; state<=ITER
  init qPrev,Q,Q_m1; cnt <= 6
 (*@\stbadge{ITER}@*) // 6 full cycles (2 digits) + 1 half
  (*@\fdivmark{Rsrt_top}@*)(q1,q2) (*@$\leftarrow$@*) DigitSelect(ws,wc,qPrev,mB)
  (ws,wc) <= CSA_Step(ws,wc,mB,qPrev,q1,q2,cnt==0)(*@\fdivmark{Rsrt_maxr}@*)
  (Q,Q_m1) <= OTFC_Step(Q,Q_m1,qPrev,q1,q2,cnt==0)
  qPrev<=q2; cnt<=cnt-1; if !cnt: state<=POST_0(*@\fdivmark{Rsrt_bot}@*)
 (*@\stbadge{POST\_0}@*)
  (*@\fdivmark{Rieee_top}@*)rem (*@$\leftarrow$@*) signed(ws+wc); qFinal (*@$\leftarrow$@*) (rem<0)?Q_m1:Q(*@\fdivmark{Rieee_maxr}@*)
  rnd (*@$\leftarrow$@*) qFinal[0]; stk (*@$\leftarrow$@*) (rem != 0)
  up (*@$\leftarrow$@*) rnd & (stk | qFinal[1])
  (cout,mant) (*@$\leftarrow$@*) qFinal[23:1] + up(*@\fdivmark{Rieee_bot}@*)
  result <= pack(sZ,eZ+cout,mant); state<=IDLE
\end{lstlisting}
\end{tcolorbox}
\end{minipage}
\end{tabular}

\begin{tikzpicture}[remember picture,overlay,>=Stealth,line width=0.55pt]
\coordinate (L_nw) at ($(L_top)+(-1.5pt,6.2pt)$);
\coordinate (L_se) at ($(L_bot)+(2pt,-2.2pt)$);
\fill[irIEEEbd,fill opacity=0.09,rounded corners=2pt] (L_nw) rectangle (L_se);
\draw[irIEEEbd!45,line width=0.4pt,rounded corners=2pt] (L_nw) rectangle (L_se);
\draw[irIEEEbd,line width=1.1pt] ($(L_nw -| L_se)+(0,-1pt)$) -- ($(L_se)+(0,1pt)$);
\coordinate (L_stage_east) at ($(L_nw -| L_se)!0.5!(L_se)$);

\coordinate (Rsrt_nw) at ($(Rsrt_top)+(-1.5pt,5.4pt)$);
\coordinate (Rsrt_se) at ($(Rsrt_maxr |- Rsrt_bot)+(2pt,-1.8pt)$);
\fill[irSRTbd,fill opacity=0.09,rounded corners=2pt] (Rsrt_nw) rectangle (Rsrt_se);
\draw[irSRTbd!45,line width=0.4pt,rounded corners=2pt] (Rsrt_nw) rectangle (Rsrt_se);
\draw[irSRTbd,line width=1.1pt] ($(Rsrt_nw)+(0,-1pt)$) -- ($(Rsrt_nw |- Rsrt_se)+(0,1pt)$);
\coordinate (Rsrt_stage_west) at ($(Rsrt_nw)!0.5!(Rsrt_nw |- Rsrt_se)$);

\coordinate (Rieee_nw) at ($(Rieee_top)+(-1.5pt,5.4pt)$);
\coordinate (Rieee_se) at ($(Rieee_maxr |- Rieee_bot)+(2pt,-1.8pt)$);
\fill[irIEEEbd,fill opacity=0.09,rounded corners=2pt] (Rieee_nw) rectangle (Rieee_se);
\draw[irIEEEbd!45,line width=0.4pt,rounded corners=2pt] (Rieee_nw) rectangle (Rieee_se);
\draw[irIEEEbd,line width=1.1pt] ($(Rieee_nw)+(0,-1pt)$) -- ($(Rieee_nw |- Rieee_se)+(0,1pt)$);
\coordinate (Rieee_stage_west) at ($(Rieee_nw)!0.5!(Rieee_nw |- Rieee_se)$);

\draw[irIEEEbd,->] (L_stage_east) -- ++(6pt,0) |- (I754.west);
\draw[irSRTbd,->]  (Rsrt_stage_west) -- ++(-6pt,0) |- (ISRT.east);
\draw[irIEEEbd,->] (Rieee_stage_west) -- ++(-6pt,0) |- (I754.east);
\draw[bordergray,->,>=Stealth,line width=0.55pt] (I754.south) |-
  node[pos=0.75,fill=white,inner sep=1.2pt,
       font=\fontsize{5.4}{6.2}\selectfont\sffamily\bfseries\itshape,
       text=black!75] {composes to}
  (boxD.east);
\end{tikzpicture}
\end{minipage}
\end{lrbox}

\resizebox{\textwidth}{!}{\usebox{\fdivFullBox}}

\par\vspace{3pt}
\begin{minipage}{\textwidth}
\fontsize{6}{7.2}\selectfont\sffamily
\textbf{Notation:} Normal-path RNE logic (corner cases/flags omitted). $\div$: floor div; $\leftarrow$/\texttt{<=}: comb./clocked update; \texttt{(ws,wc)}/\texttt{(Q,Q\_m1)}: CSA/quotient registers; \texttt{mA}/\texttt{mB}/\texttt{mZ} include hidden bit.
\end{minipage}

\caption{Semantic decomposition of \texttt{FDIV} via Speculated IR models: \textbf{(a)}~SoftFloat REF; \textbf{(b)}~Fudian radix-4 DUV; \textbf{(c)}~$\mathrm{IR}_{\mathrm{SRT}}$ (purple, SRT integer contract) and $\mathrm{IR}_{754}$ (blue, exact division and IEEE~754 rounding); \textbf{(d)}~full-design equivalence for all inputs, rounding modes.}
\label{fig:gap}
\end{figure*}

%% file: figs/fig-falseloop.tex
\begin{figure}[!htbp]
\centering
\resizebox{\linewidth}{!}{
\begin{tikzpicture}[
  font=\scriptsize,
  box/.style={draw=black!45, rounded corners=2pt, align=center,
    inner sep=2pt, line width=0.5pt},
  ar/.style={-{Latex[length=1.0mm]}, draw=black!65, line width=0.5pt},
  falsear/.style={ar, draw=red!65!black, dashed},
  wire/.style={ar, draw=blue!55!black},
  heading/.style={font=\scriptsize\bfseries, anchor=base,
    text height=1.5ex, text depth=0.25ex},
  note/.style={font=\tiny, text=black!65, align=center,
    fill=white, inner sep=1pt},
  code/.style={box, align=left, inner sep=3.5pt, font=\fontsize{6}{7.2}\selectfont}
]

\node[heading] at (1.35,0) {(a) Monolithic Call};
\node[heading] at (4.30,0) {(b) True Dependencies};
\node[heading] at (7.35,0) {(c) Per-Output Split};

\draw[black!15] (2.75,-0.35) -- (2.75,-2.75);
\draw[black!15] (5.82,-0.35) -- (5.82,-2.75);

\node[box, fill=red!4, minimum width=1.1cm, minimum height=10mm]
  (whole) at (1.35,-1.6) {\texttt{add}};
\node (a0) at (0.20,-1.35) {$a$};
\node (b0) at (0.20,-1.85) {$b$};
\node (x0) at (2.50,-1.35) {$x$};
\node (y0) at (2.50,-1.85) {$y$};
\draw[falsear] (a0.east) -- ([yshift=2.5mm]whole.west);
\draw[ar] (b0.east) -- ([yshift=-2.5mm]whole.west);
\draw[falsear] ([yshift=2.5mm]whole.east) -- (x0.west);
\draw[ar] ([yshift=-2.5mm]whole.east) -- (y0.west);
\draw[wire] (x0.north) -- (2.50,-0.55) -- (0.20,-0.55) -- (a0.north);
\node[note, text=blue!55!black] at (1.35,-0.55) {$a:=x$};

\draw[box, dashed] (3.32,-0.85) rectangle (5.28,-2.65);
\node[note] at (4.30,-0.85) {\texttt{add}};
\node[box, fill=green!5, minimum width=7mm, minimum height=4mm]
  (const) at (3.92,-1.2) {\texttt{0x11}};
\node[circle, draw=black!65, fill=blue!4, minimum size=4mm,
      inner sep=0pt, font=\small] (sum) at (4.58,-2.1) {$+$};
\node (a1) at (3.08,-1.85) {$a$};
\node (b1) at (3.08,-2.35) {$b$};
\node (x1) at (5.52,-1.2) {$x$};
\node (y1) at (5.52,-2.1) {$y$};
\draw[ar] (const.east) -- (x1.west);
\draw[ar] (sum.east) -- (y1.west);
\draw[ar] (a1.east) -- (4.08,-1.85) |- (sum.150);
\draw[ar] (b1.east) -- (4.08,-2.35) |- (sum.210);
\draw[wire] (x1.north) -- (5.52,-0.55) -- (3.08,-0.55) -- (a1.north);
\node[note, text=blue!55!black] at (4.30,-0.55) {$a:=x$};

\node[code, fill=green!5, anchor=west] (ux) at (6.35,-0.92)
  {\texttt{let x\_out :=}\\\texttt{~~(add.outputs \{\}).x}};
\node[code, fill=blue!4, anchor=west] (uy) at (6.35,-2.22)
  {\texttt{let y\_out :=}\\\texttt{~~(add.outputs}\\\texttt{~~~\{a:=x\_out, b:=b\}).y}};
\node (input) at ([xshift=-3.5mm]uy.west) {$b$};
\draw[ar] (input.east) -- (uy.west);
\draw[wire] (ux.south) -- node[note, right, xshift=1pt] {$a:=x_{\text{out}}$} (ux.south |- uy.north);

\end{tikzpicture}
}
\caption{Eliminating false combinational loops via per-output evaluation. (a) Monolithic invocation creates a false $a \to x$ cycle; (b) Inside \texttt{add}, $x$ is constant and $y = a + b$, with no path from $a$ to $x$; (c) Per-output evaluation computes $x_{\text{out}}$ first (defaulting unused inputs to $0$) and then $y_{\text{out}}$ with $a = x_{\text{out}}$.}
\label{fig:falseloop}
\end{figure}
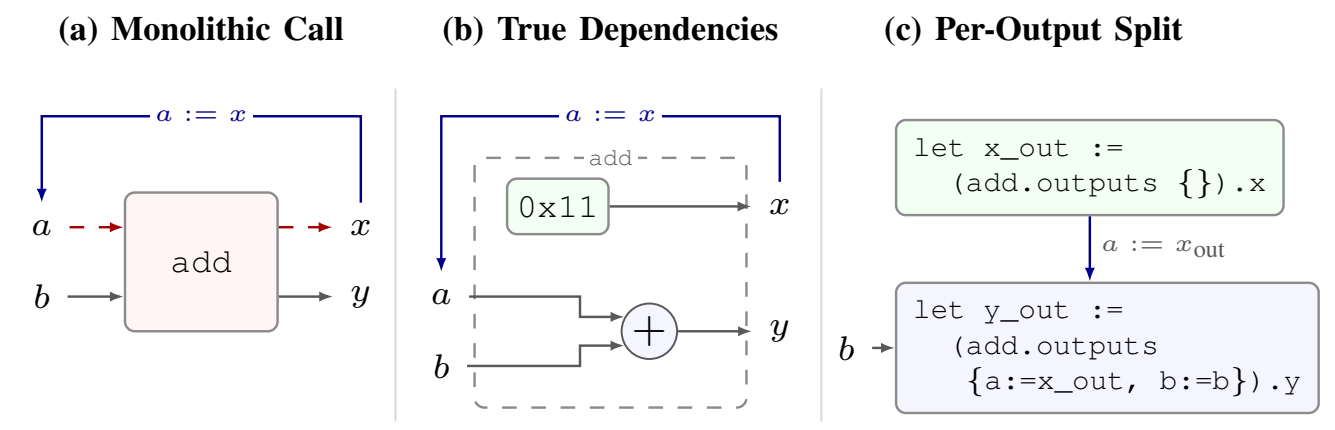

%% file: sections/4_evaluations.tex
\section{Evaluation}
\label{sec:eval}

\subsection{Experimental Setup}

\paragraph{Benchmark Circuits}
We test our system on three open-source hardware benchmarks covering both datapaths and control logic:
(1)~\textbf{Floating-Point Units:} single-precision arithmetic modules from the XiangShan~\cite{xiangshan} Fudian FPU, including \texttt{fadd}, \texttt{fmul}, \texttt{fdiv}, and \texttt{fma}, checked against Berkeley SoftFloat~\cite{softfloat};
(2)~\textbf{Convolution Unit:} the 8-bit integer (\texttt{int8}) \texttt{CMAC} datapath from NVIDIA NVDLA~\cite{nvdla}, checked against its official C++ reference model; and
(3)~\textbf{Processor Core:} the multi-cycle RISC-V core \texttt{PicoRV32}~\cite{picorv32} executing the 38-instruction RV32I base ISA, checked against the official Sail~\cite{sail} ISA spec.

\paragraph{Baselines}
We compare TheoremV against three baselines:
(1)~\textbf{HEC:} the commercial High-Level Equivalence Checking tool in X-EPIC GalaxFV, which checks C/C++-to-Verilog equivalence directly;
(2)~\textbf{Rtl2lean~\cite{rtl2lean-paper}:} its PySlang AST-based RTL-to-Lean translator plugged into our proving pipeline to test frontend representations, since its original workflow mainly checks properties rather than full equivalence between the RTL and reference models; and
(3)~\textbf{Ours w/o Speculated IR:} an ablation that removes the explicit IR specification from the workflow and asks the agent to prove equivalence directly between the translated DUV and REF using the same LLM setup.

\paragraph{Environment and Setup}
All proofs are built and checked in Lean~v4.29.0 with Mathlib on a Linux server with Intel Xeon Gold 6254 CPUs. Both TheoremV and the LLM baselines run on Claude Code~v2.1.319 with DeepSeek-V4.1-Flash-1017, using up to five parallel subagents. We cap each Lean build at 1\,h and set a 24\,h timeout per task for all methods, extending it to 48\,h for the complex \texttt{fma} benchmark.

\paragraph{Evaluation Metrics}

To measure system efficiency, we report \textbf{Time} and \textbf{Tokens}. Time is the total time when at least one agent is actively generating responses or running tools. Tokens measures the total input and output tokens used by all agents.

\subsection{Main Results}
\label{sec:eval:scale}

Tables~\ref{tab:baseline} and~\ref{tab:proof_scale} report verification performance, resource overheads, and proof scales across all benchmarks. TheoremV successfully closes all formal proofs from RTL to the golden reference on every circuit.

\begin{table}[t]
\centering
\setlength{\abovecaptionskip}{2pt}
\setlength{\belowcaptionskip}{4pt}
\caption{Verification capabilities and resource overhead across benchmarks. \texttt{TO} denotes timeout (exceeding the respective wall-clock limits); \texttt{Unsup.} denotes unsupported.}
\label{tab:baseline}
\scriptsize
\setlength{\tabcolsep}{3pt}
\renewcommand{\arraystretch}{1.08}
\setlength{\aboverulesep}{1.5pt}
\setlength{\belowrulesep}{1.5pt}
\resizebox{\columnwidth}{!}{
\begin{tabular}{@{}l c c cc cc@{}}
\toprule
\multirow{2}{*}{\textbf{Benchmark}} 
  & \textbf{HEC} 
  & \textbf{Rtl2lean}
  & \multicolumn{2}{c}{\makecell{\textbf{Ours (w/o}\\\textbf{Speculated IR)}}} 
  & \multicolumn{2}{c}{\textbf{Ours}} \\
\cmidrule(lr){2-2} \cmidrule(lr){3-3} \cmidrule(lr){4-5} \cmidrule(lr){6-7}
& Time 
  & Time 
  & Time & Tokens 
  & Time & Tokens \\
\midrule
\texttt{fadd} 
  & \textbf{25s}
  & Lean syntax error
  & 8h\,51m & 7.62M 
  & 8h\,03m & \textbf{2.39M} \\
\texttt{fmul} 
  & TO ($>24$h) 
  & Build TO ($>1$h)
  & 8h\,34m & \textbf{6.12M} 
  & \textbf{7h\,58m} & 6.94M \\
\texttt{fdiv} 
  & TO ($>24$h) 
  & Build TO ($>1$h)
  & 20h\,34m & 13.04M 
  & \textbf{10h\,46m} & \textbf{12.42M} \\
\texttt{fma} 
  & TO ($>24$h) 
  & Build TO ($>1$h)
  & \multicolumn{2}{c}{TO ($>48$h)} 
  & \textbf{29h\,39m} & \textbf{63.61M} \\
\midrule
\texttt{CMAC} 
  & TO ($>24$h) 
  & Unsup.\ Verilog syntax
  & 10h\,01m & \textbf{7.25M} 
  & \textbf{6h\,48m} & 9.31M \\
\midrule
\texttt{PicoRV32} 
  & Unsup.
  & TO ($>24$h) 
  & \multicolumn{2}{c}{TO ($>24$h)} 
  & \textbf{12h\,59m} & \textbf{19.05M} \\
\bottomrule
\end{tabular}
}
\end{table}

\begin{table}[t]
    \centering
    \scriptsize
    \setlength{\abovecaptionskip}{2pt}
    \setlength{\belowcaptionskip}{4pt}
    \caption{Scale of Lean models and proofs across benchmarks.}
    \label{tab:proof_scale}
    \setlength{\tabcolsep}{4.5pt}
    \renewcommand{\arraystretch}{1.05}
    \setlength{\aboverulesep}{1.5pt}
    \setlength{\belowrulesep}{1.5pt}
    \begin{tabular}{@{}lrrrr@{}}
    \toprule
    \textbf{Benchmark} & \textbf{DUV LoC} & \textbf{REF LoC} & \textbf{Proof LoC} & \textbf{Theorems} \\
    \midrule
    \texttt{fadd}     & 762      & 1{,}438  & 1{,}826   & 74      \\
    \texttt{fmul}     & 1{,}318  & 1{,}238  & 21{,}782  & 1{,}355 \\
    \texttt{fdiv}     & 2{,}233  & 1{,}337  & 35{,}026  & 1{,}631 \\
    \texttt{fma}      & 2{,}186  & 5{,}367  & 170{,}923 & 5{,}605 \\
    \texttt{CMAC}     & 35{,}320 & 561      & 121{,}175 & 15{,}521 \\
    \texttt{PicoRV32} & 1{,}240  & 56{,}482 & 77{,}580  & 2{,}609 \\
    \bottomrule
    \end{tabular}
\end{table}

\paragraph{TheoremV Surpasses HEC in Scalability and Expressiveness}
Although HEC verifies the lightweight, linear \texttt{fadd} datapath in 25\,s, it times out after 24\,h on all remaining arithmetic benchmarks due to SAT/SMT clause blowup and cannot encode ISA-to-RTL refinement for \texttt{PicoRV32}. In contrast, TheoremV avoids circuit unrolling by combining algebraic abstraction and inductive reasoning in Lean~4, closing end-to-end proofs for \texttt{fmul}, \texttt{fdiv}, \texttt{fma}, and \texttt{CMAC} in 6h\,48m to 29h\,39m across all inputs. Furthermore, Lean~4's expressive logic naturally captures cycle-to-instruction refinement relations, enabling TheoremV to check \texttt{PicoRV32} across the 38-instruction RV32I base set against Sail in 12h\,59m and uncover a real \texttt{FENCE} bug (Section~\ref{sec:bugs}).

\paragraph{Hierarchy-Preserving, Executable Compilation Is Necessary for Full Proofs}
Under the same agent setup, Rtl2lean fails to finish whole-design equivalence proofs due to three frontend limits. First, its direct AST translator covers a narrower SystemVerilog subset, failing to import \texttt{fadd} and \texttt{CMAC}. Second, it flattens the whole design globally, causing Lean builds to exceed the 1\,h cap. Third, on \texttt{PicoRV32}, its Lean model repeatedly unfolds combinational logic, which blocks proof convergence. These results show that hierarchy-preserving, executable RTL-to-Lean compilation with broad syntax coverage is necessary for finishing full proofs.

\paragraph{Speculated IRs Scale and Speed Up Proving}
Speculated IRs cut runtime by up to 47.6\% on arithmetic units and enable full proofs for \texttt{fma} and \texttt{PicoRV32}. Without these speculatively generated intermediate models, agents must link raw DUV and REF code directly with ad-hoc lemmas, which slows down proofs and limits scale. On math units, Speculated IRs split big proofs into smaller parallel goals, letting the complex \texttt{fma} unit finish within the 48\,h limit. On \texttt{PicoRV32}, proving multi-cycle RTL directly against single-step Sail specs often stalls and times out; adding a single-cycle IR folds multi-cycle steps into clean instruction bounds, resolving this gap. While Speculated IRs raise token usage by 13.5\% on \texttt{fmul}, this reflects a clear trade-off. Without IRs, the baseline compares both models directly using long-timeout SAT calls like \texttt{bv\_decide}, keeping code short but prone to timeouts. With IRs, agents separate both sides and prove each against a pure math spec via small lemmas and short SAT calls, growing proof code by 4.4$\times$. Writing these explicit proofs uses slightly more output tokens, yet it prevents solver stalls and speeds up overall runtime.

\paragraph{Hierarchy Preservation Enables Lemma Reuse and Cuts Proof Overhead}
To evaluate hierarchy preservation, we flattened the \texttt{fmul} RTL before Lean translation under identical settings. While it still compiles, proof time increases from 7h\,58m to 13h\,20m.
Preserving hierarchy keeps clear boundaries around compressor submodules so the agent can split and reuse local lemmas, whereas flattening merges all internal logic into a single large scope, increasing both proof search and compilation overhead.

\paragraph{TheoremV Scales Formal Proofs Far Beyond Manual Effort}
Table~\ref{tab:proof_scale} shows sharp size asymmetries between DUV and REF (35{,}320 vs.\ 561 LoC on \texttt{CMAC}, and 1{,}240 vs.\ 56{,}482 LoC on \texttt{PicoRV32}), which reflect wide semantic differences between the two implementations and therefore make direct alignment difficult. Yet TheoremV still automatically builds 1{,}826 to 170{,}923 LoC of Lean proofs (5{,}605 theorems on \texttt{fma}) and up to 15{,}521 theorems (\texttt{CMAC}), far beyond practical manual effort.
\subsection{Case Study}
\label{sec:bugs}

\paragraph{Uncovering a Real Spec-Violating Hardware Bug in PicoRV32}
The RISC-V specification requires processors to ignore the reserved \texttt{rs1} and \texttt{rd} fields in the \texttt{FENCE} instruction. However, during proof search, \texttt{PicoRV32} failed the subgoal ensuring that the register file stays unchanged when executing \texttt{FENCE}. We built and verified the following counterexample sequence:
\begin{quote}
\ttfamily
addi x1, x0, 5;  fence (rd=1);
\end{quote}

\input{figs/fig-dutbug}

As illustrated in Figure~\ref{fig:dutbug}, after execution, Sail ignores \texttt{rd} and keeps \texttt{x1 = 5}, whereas \texttt{PicoRV32} falls through to its default write-back path and executes \texttt{regs[rd] <= 0}, wrongly clearing \texttt{x1} to zero.

\paragraph{Testing and Proof Feedback Repairs Flawed Speculated IRs}
Beyond finding real hardware bugs, our system also catches and fixes mistakes in its IRs during proof search. On \texttt{fma}, the agent's initial IEEE-754 model had four edge-case bugs, such as wrongly setting the underflow flag on exact subnormal results and missing the invalid flag when $0 \times \infty$ appears with a NaN. Running one million differential tests caught and fixed these bugs before formal proving began, avoiding wasted work on false goals. On \texttt{fdiv}, subagents found counterexamples during local checks and sent flawed SRT division specs back to the main agent for fixes, correcting wrong remainder bounds and selection factors. This feedback loop keeps the intermediate models accurate and helps finish the full proof smoothly.

\subsection{Limitations and Future Work}
TheoremV targets synchronous single-clock synthesizable RTL, covering standard datapaths and processor cores. Extending support to multi-clock or gated-clock domains, tri-state/\texttt{inout} interfaces, and \texttt{x}/\texttt{z} semantics is left for future work. Also, while Speculated IRs enable proving complex logic, they can add token overhead on tasks where direct proving already succeeds; we plan to explore adaptive strategies that balance layered IR construction with module reuse to reduce proof cost on larger designs.

%% file: figs/fig-dutbug.tex
\begin{figure}[!htbp]
\centering
\resizebox{\linewidth}{!}{
\begin{tikzpicture}[
  font=\scriptsize,
  box/.style={draw=black!45, rounded corners=2pt, align=center,
    inner sep=2.5pt, line width=0.5pt},
  ar/.style={-{Latex[length=1.0mm]}, draw=black!65, line width=0.5pt},
  heading/.style={font=\scriptsize\bfseries, anchor=base,
    text height=1.5ex, text depth=0.25ex},
  note/.style={font=\fontsize{5.2}{6.2}\selectfont, text=black!65, align=center,
    fill=white, inner sep=1.2pt},
  codebox/.style={box, font=\fontsize{6}{7.2}\selectfont\ttfamily},
  modelbox/.style={box, fill=blue!4, font=\fontsize{6}{7.2}\selectfont},
  stateok/.style={box, fill=green!5, font=\fontsize{6}{7.2}\selectfont\ttfamily},
  stateerr/.style={box, fill=red!4, font=\fontsize{6}{7.2}\selectfont\ttfamily}
]

\node[codebox, fill=black!2, minimum width=1.85cm, minimum height=8.5mm]
  (prog) at (1.25, -0.55) {addi x1, x0, 5\\fence (rd=1)};

\node[modelbox, minimum width=1.60cm, minimum height=5.0mm]
  (sail) at (4.10, 0.00) {\textbf{Sail}};
\node[modelbox, minimum width=1.60cm, minimum height=5.0mm]
  (pico) at (4.10, -1.10) {\textbf{PicoRV32}};

\node[stateok, minimum width=1.45cm, minimum height=8.0mm]
  (sailrf) at (7.55, 0.00) {{\color{black!50}x0 = 0}\\{\bfseries\color{green!40!black}x1 = 5}};
\node[stateerr, minimum width=1.45cm, minimum height=8.0mm]
  (picorf) at (7.55, -1.10) {{\color{black!50}x0 = 0}\\{\bfseries\color{red!65!black}x1 = 0}};

\coordinate (fork) at (2.70, -0.55);
\draw[draw=black!65, line width=0.5pt] (prog.east) -- (fork);
\draw[ar] (fork) |- (sail.west);
\draw[ar] (fork) |- (pico.west);

\draw[ar] (sail.east) --
  node[note, above=2pt] {Ignores \texttt{rd}}
  (sailrf.west);
\draw[ar] (pico.east) --
  node[note, text=red!65!black, above=2pt] {Erroneous\\writeback}
  (picorf.west);

\node[heading] at (1.25, -1.82) {Input Sequence};
\node[heading] at (4.10, -1.82) {Execution Model};
\node[heading] at (7.55, -1.82) {Post-Execution State};

\end{tikzpicture}
}
\caption{PicoRV32 \texttt{fence} writeback bug: Sail ignores reserved \texttt{rd} (\texttt{x1 = 5}), whereas PicoRV32 erroneously writes back (\texttt{x1 = 0}).}
\label{fig:dutbug}
\end{figure}
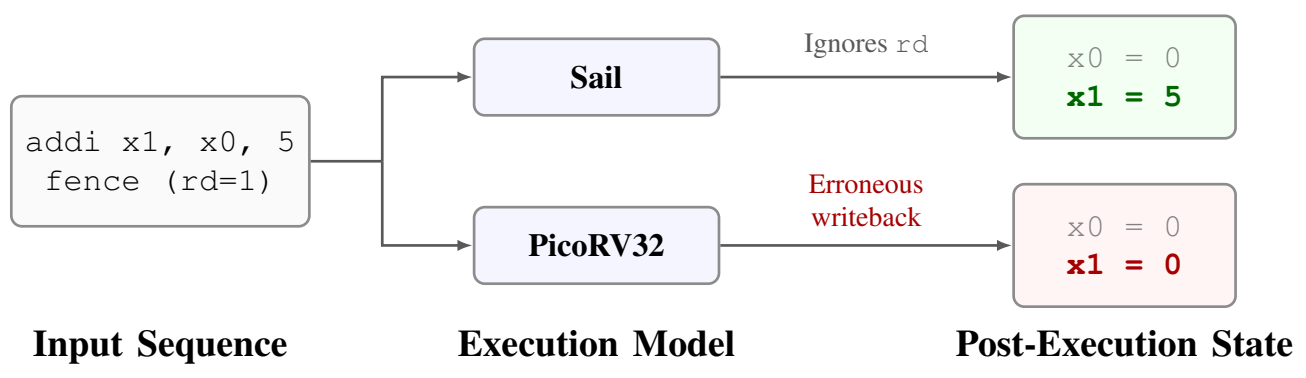

%% file: sections/99_conclusion.tex
\section{Conclusion}
\label{sec:conclusion}

In this work, we introduced \textbf{TheoremV}, showing that agentic theorem proving can automate end-to-end functional equivalence verification for complex, large-scale RTL without manual proof libraries. By pairing a \textit{Speculated IR}-based agent workflow with a hierarchy-preserving, executable RTL-to-Lean compiler, TheoremV bridges both the abstraction gap and the representation gap within a unified Lean~4 flow. Our evaluation on \texttt{Fudian}, \texttt{CMAC}, and \texttt{PicoRV32} demonstrates that this approach scales to formal proofs of up to 170{,}923 lines of proof code on circuits beyond the reach of commercial equivalence checkers and prior provers, while uncovering a real \texttt{FENCE} spec violation in the processor core. Future work will extend support to multi-clock circuits and adaptive IR reuse.

%% file: sections/100_AI.tex
\section*{Content Generated by AI}
Large language models were used to assist with preliminary literature search and language polishing of the manuscript. All technical claims, experiments, and conclusions were completed and verified by the authors, who take full responsibility for the content of this paper.

%% file: references.bib
@inproceedings{legoprover,
title={{LEGO}-Prover: Neural Theorem Proving with Growing Libraries},
author={Haiming Wang and Huajian Xin and Chuanyang Zheng and Zhengying Liu and Qingxing Cao and Yinya Huang and Jing Xiong and Han Shi and Enze Xie and Jian Yin and Zhenguo Li and Xiaodan Liang},
booktitle={The Twelfth International Conference on Learning Representations},
year={2024},
url={https://openreview.net/forum?id=3f5PALef5B}
}

@inproceedings{dsp,
title={Draft, Sketch, and Prove: Guiding Formal Theorem Provers with Informal Proofs},
author={Albert Qiaochu Jiang and Sean Welleck and Jin Peng Zhou and Timothee Lacroix and Jiacheng Liu and Wenda Li and Mateja Jamnik and Guillaume Lample and Yuhuai Wu},
booktitle={The Eleventh International Conference on Learning Representations },
year={2023},
url={https://openreview.net/forum?id=SMa9EAovKMC}
}

@misc{yosys,
  author       = {Claire Wolf},
  title        = {{Yosys} Open {SYnthesis} Suite},
  howpublished = {\url{https://yosyshq.net/yosys/}}
}

@article{sail,
  title={ISA semantics for ARMv8-a, RISC-v, and CHERI-MIPS},
  author={Armstrong, Alasdair and Bauereiss, Thomas and Campbell, Brian and Reid, Alastair and Gray, Kathryn E and Norton, Robert M and Mundkur, Prashanth and Wassell, Mark and French, Jon and Pulte, Christopher and others},
  journal={Proceedings of the ACM on Programming Languages},
  volume={3},
  number={POPL},
  pages={1--31},
  year={2019},
  publisher={ACM, New York, NY, USA}
}

@inproceedings{xiangshan,
  title={Towards developing high performance RISC-V processors using agile methodology},
  author={Xu, Yinan and Yu, Zihao and Tang, Dan and Chen, Guokai and Chen, Lu and Gou, Lingrui and Jin, Yue and Li, Qianruo and Li, Xin and Li, Zuojun and others},
  booktitle={2022 55th IEEE/ACM International Symposium on Microarchitecture (MICRO)},
  pages={1178--1199},
  year={2022},
  organization={IEEE}
}

@article{kami,
  title={Kami: a platform for high-level parametric hardware specification and its modular verification},
  author={Choi, Joonwon and Vijayaraghavan, Muralidaran and Sherman, Benjamin and Chlipala, Adam and Arvind},
  journal={Proceedings of the ACM on Programming Languages},
  volume={1},
  number={ICFP},
  pages={1--30},
  year={2017},
  publisher={ACM, New York, NY, USA}
}

@inproceedings{harrison-fdiv,
  title={Formal verification of IA-64 division algorithms},
  author={Harrison, John},
  booktitle={International Conference on Theorem Proving in Higher Order Logics},
  pages={233--251},
  year={2000},
  organization={Springer}
}

@inproceedings{kaivola,
  title={Formal verification of the {Pentium} 4 floating-point multiplier},
  author={Kaivola, Roope and Narasimhan, Naren},
  booktitle={Proceedings 2002 Design, Automation and Test in Europe Conference and Exhibition},
  pages={20--27},
  year={2002},
  organization={IEEE}
}

@inproceedings{flocq,
  title={Flocq: A unified library for proving floating-point algorithms in Coq},
  author={Boldo, Sylvie and Melquiond, Guillaume},
  booktitle={2011 IEEE 20th Symposium on Computer Arithmetic},
  pages={243--252},
  year={2011},
  organization={IEEE}
}

@article{deepseekprover,
  title={Deepseek-prover: Advancing theorem proving in llms through large-scale synthetic data},
  author={Xin, Huajian and Guo, Daya and Shao, Zhihong and Ren, Zhizhou and Zhu, Qihao and Liu, Bo and Ruan, Chong and Li, Wenda and Liang, Xiaodan},
  journal={arXiv preprint arXiv:2405.14333},
  year={2024}
}

@article{alphaproof,
  title={Olympiad-level formal mathematical reasoning with reinforcement learning},
  author={Hubert, Thomas and Mehta, Rishi and Sartran, Laurent and Horv{\'a}th, Mikl{\'o}s Z and {\v{Z}}u{\v{z}}i{\'c}, Goran and Wieser, Eric and Huang, Aja and Schrittwieser, Julian and Schroecker, Yannick and Masoom, Hussain and others},
  journal={Nature},
  volume={651},
  number={8106},
  pages={607--613},
  year={2026},
  publisher={Nature Publishing Group UK, London}
}

@article{alphaproof-nexus,
  title={Advancing mathematics research with AI-driven formal proof search},
  author={Tsoukalas, George and Kovsharov, Anton and Shirobokov, Sergey and Surina, Anja and Firsching, Moritz and B{\'e}rczi, Gergely and Ruiz, Francisco JR and Suggala, Arun and Wagner, Adam Zsolt and Wieser, Eric and others},
  journal={Science},
  volume={394},
  number={6820},
  pages={234--239},
  year={2026},
  publisher={American Association for the Advancement of Science}
}

@article{cktformalizer,
  title={CktFormalizer: Autoformalization of Natural Language into Circuit Representations},
  author={Xiong, Jing and Han, Qi and Ding, Chenchen and Xiao, He and Su, Zunhai and Tao, Chaofan and Wong, Ngai},
  journal={arXiv preprint arXiv:2605.07782},
  year={2026}
}

@inproceedings{chicala,
  title={Formally Verifying Arithmetic Chisel Designs for All Bit Widths at Once},
  author={Feng, Weizhi and Liu, Yicheng and Liu, Jiaxiang and Jansen, David N and Zhang, Lijun and Wu, Zhilin},
  booktitle={Proceedings of the 61st ACM/IEEE Design Automation Conference},
  pages={1--6},
  year={2024}
}

@article{ooo-multiprocessor,
  title={Formal Verification of an Out-of-Order Multiprocessor against an In-Order Weak-Memory ISA},
  author={Lee, Janggun and Kang, Jeehoon},
  journal={arXiv preprint arXiv:2607.18727},
  year={2026}
}

@inproceedings{loow2022,
  author={Lööw, Andreas},
  booktitle={2022 Formal Methods in Computer-Aided Design (FMCAD)}, 
  title={Reconciling Verified-Circuit Development and Verilog Development}, 
  year={2022},
  volume={},
  number={},
  pages={1-10},
  doi={10.34727/2022/isbn.978-3-85448-053-2_15}}

@article{rtl2lean-paper,
  title={Rtl2lean: Automated RTL-to-Lean Translation with Hierarchical Theorem Generation and Lemma Reuse},
  author={Lyu, Hongqin and Dong, Junxing and Wang, Yonghao and Chao, Zhiteng and Wang, Tiancheng and Li, Huawei},
  journal={arXiv preprint arXiv:2607.16855},
  year={2026}
}

@article{circuitprover,
  title={CircuitProver: Agentic Lean 4 Theorem Proving with Reusable Circuit Proof Library for Hardware Verification},
  author={Yang, Ziyi and Fang, Wenji and Chen, Chen and Xie, Zhiyao and Zhang, Hongce},
  journal={arXiv preprint arXiv:2607.27259},
  year={2026}
}

@article{revamping-verilog,
  title={Revamping Verilog semantics for foundational verification},
  author={Choi, Joonwon and Kim, Jaewoo and Kang, Jeehoon},
  journal={Proceedings of the ACM on Programming Languages},
  volume={9},
  number={OOPSLA2},
  pages={950--977},
  year={2025},
  publisher={ACM, New York, NY, USA}
}

@article{cill,
  title={CIll: CTI-Guided Invariant Generation via LLMs for Model Checking},
  author={Su, Yuheng and Bu, Tianjun and Yang, Qiusong and Ci, Yiwei and Tian, Enyuan},
  journal={arXiv preprint arXiv:2602.23389},
  year={2026}
}

@inproceedings{infiniverif,
title={InfiniVerif: Closing the Trust Gap in Agentic Hardware Design via Fully Automated Formal Verification},
author={Yiran Xia and Qi Liu and Bangyan Wang and Yuan Xie},
booktitle={AI for Chip Design - NeurIPS 2026 Workshop},
year={2026},
url={https://openreview.net/forum?id=fk53pcimJT}
}

@article{revisiting-verilogeval,
  title={Revisiting verilogeval: A year of improvements in large-language models for hardware code generation},
  author={Pinckney, Nathaniel and Batten, Christopher and Liu, Mingjie and Ren, Haoxing and Khailany, Brucek},
  journal={ACM Transactions on Design Automation of Electronic Systems},
  volume={30},
  number={6},
  pages={1--20},
  year={2025},
  publisher={ACM, New York, NY}
}

@misc{softfloat,
  author       = {John R. Hauser},
  title        = {Berkeley {SoftFloat} Release 3e},
  howpublished = {\url{https://www.jhauser.us/arithmetic/SoftFloat.html}},
  year         = {2018},
  month        = jan
}

@misc{riscvformal,
  author       = {{YosysHQ}},
  title        = {{riscv-formal}: {RISC-V} Formal Verification Framework},
  howpublished = {\url{https://github.com/YosysHQ/riscv-formal}},
  note         = {Accessed: 2026-11-16}
}

@misc{picorv32,
  author       = {Claire Xenia Wolf},
  title        = {{PicoRV32} -- A Size-Optimized {RISC-V} {CPU}},
  howpublished = {\url{https://github.com/YosysHQ/picorv32}},
  note         = {Accessed: 2026-11-16}
}

@misc{sparkle,
  author       = {Junji Hashimoto},
  title        = {Sparkle: A Type-Safe, Formally Verifiable {HDL} Compiler in {Lean} 4},
  howpublished = {\url{https://github.com/Verilean/sparkle}},
  note         = {Accessed: 2026-11-16}
}

@misc{nvdla,
  author       = {{NVIDIA Corporation}},
  title        = {{NVIDIA Deep Learning Accelerator (NVDLA)}},
  howpublished = {\url{http://nvdla.org/}},
  year         = {2017}
}
